\documentclass[12pt]{article}
\usepackage{amssymb}
\usepackage{amsmath}
\usepackage{psfig}
\begin{document}

\title{Some consequences of a noncommutative space-time structure}
\date{}
\author{R. Vilela Mendes\thanks{%
Grupo de F\'{i}sica Matem\'{a}tica and Universidade T\'{e}cnica de Lisboa,
Complexo Interdisciplinar, Av. Gama Pinto 2, 1649-003 Lisboa, Portugal,
vilela@cii.fc.ul.pt}}
\maketitle

\begin{abstract}
The existence of a fundamental length (or fundamental time) has been
conjecture in many contexts. Here one discusses some consequences of a
fundamental constant of this type, which emerges as a consequence of
deformation-stability considerations leading to a non-commutative space-time
structure. This mathematically well defined structure is sufficiently
constrained to allow for unambiguous experimental predictions. In particular
one discusses the phase-space volume modifications and their relevance for
the calculation of the GZK sphere. Corrections to the spectrum of the
Coulomb problemb are also computed.
\end{abstract}

PACS: 13.60.-r; 03.65.Bz; 98.70.Sa

\section{Introduction}

Motivated by string theory and quantum gravity, many studies have been
performed exploring the non-commutative effects that are expected to appear
at the Planck scale. Associated to this is also the role played by a
fundamental length (or fundamental time) as a new constant of Nature.
However, in my opinion, the most satisfactory and model-independent way to
approach this problem is through deformation theory and considerations of
structural stability of the physical theories.

Indeed, the transition from non-relativistic to relativistic and from
classical to quantum mechanics, may be interpreted as the replacement of two
unstable theories by two stable ones. That is, by theories that do not
change in a qualitative manner under a small change of parameters. The
deformation parameters are $\frac{1}{c}$ (the inverse of the speed of light)
and $h$ (the Planck constant). Stability arises from the fact that the
algebraic structures are all equivalent for non-zero values of $\frac{1}{c}$
and $h$. The zero value is an isolated point corresponding to the
deformation-unstable classical theories.

A similar stability analysis of relativistic quantum mechanics \cite{Vilela1}
\cite{Vilela2} leads to a non-commutative space-time algebra $\Re _{\ell
,\infty }$ (on the tangent space) 
\begin{equation}
\begin{array}{rcl}
\lbrack M_{\mu \nu },M_{\rho \sigma }] & = & i(M_{\mu \sigma }\eta _{\nu
\rho }+M_{\nu \rho }\eta _{\mu \sigma }-M_{\nu \sigma }\eta _{\mu \rho
}-M_{\mu \rho }\eta _{\nu \sigma }) \\ 
\lbrack M_{\mu \nu },p_{\lambda }] & = & i(p_{\mu }\eta _{\nu \lambda
}-p_{\nu }\eta _{\mu \lambda }) \\ 
\lbrack M_{\mu \nu },x_{\lambda }] & = & i(x_{\mu }\eta _{\nu \lambda
}-x_{\nu }\eta _{\mu \lambda }) \\ 
\lbrack p_{\mu },p_{\nu }] & = & 0 \\ 
\lbrack x_{\mu },x_{\nu }] & = & -i\epsilon \ell ^{2}M_{\mu \nu } \\ 
\lbrack p_{\mu },x_{\nu }] & = & i\eta _{\mu \nu }\Im  \\ 
\lbrack p_{\mu },\Im ] & = & 0 \\ 
\lbrack x_{\mu },\Im ] & = & i\epsilon \ell ^{2}p_{\mu } \\ 
\lbrack M_{\mu \nu },\Im ] & = & 0
\end{array}
\label{1.1}
\end{equation}
and to two new parameters $\left( \ell ,\epsilon \right) $, $\ell $ being a
fundamental length (or fundamental time) and $\epsilon $ a sign $\left(
\epsilon =-1\text{ or }\epsilon =+1\right) $. In Eqs.(\ref{1.1}) $\eta _{\mu
\nu }=(1,-1,-1,-1)$, $c=\hbar =1$ and $\Im $ is the operator that replaces
the trivial center of the Heisenberg algebra.

The non-commutative space-time geometry arising from this algebra has been
studied \cite{Vilela3}, as well as the modification of the uncertainty
relations \cite{Eric}.

Here I will concentrate in some consequences of this non-commutative
structure which might lead to simpler experimental tests. In particular
phase-space suppression or enhancing effects will be discussed and their
relevance to the calculation of the GZK sphere as well as the corrections to
the spectrum of the Coulomb problem.

Notice that the modifications introduced on the calculation of the GZK
sphere do not arise from violation of Lorentz invariance, which is well
preserved, but from a change on the cross sections due to a phase-space
volume suppression at high energies. The phase-space suppression only occurs
if $\epsilon =+1$. If $\epsilon =-1$ there would be a phase-space enhancing.
The $\epsilon =-1$ and $\epsilon =+1$ cases are also quite different as far
as the spectrum of the space-time coordinates is concerned. In the first
case it is a space coordinate that has discrete spectrum, whereas the time
spectrum is continuous. In the second, it is time that is discrete, space
always having continuous spectrum.

\section{Phase-space effects arising from non-commutativity}

Here we see that depending on the sign of $\epsilon $, the available phase
space volume at high momentum contracts or expands. First, this will be
shown in the framework of a full representation of the algebra and then, to
obtain a simple analytical estimate of the effect, a simpler representation
of a subalgebra will be used.

Let 
\begin{equation}
\begin{array}{lll}
p_{\mu } & = & i\frac{\partial }{\partial \xi ^{\mu }} \\ 
\Im  & = & i\ell \frac{\partial }{\partial \xi ^{4}} \\ 
x_{\mu } & = & i\ell (\xi _{\mu }\frac{\partial }{\partial \xi ^{4}}%
-\epsilon \xi ^{4}\frac{\partial }{\partial \xi ^{\mu }}) \\ 
M_{\mu \nu } & = & i(\xi _{\mu }\frac{\partial }{\partial \xi ^{\nu }}-\xi
_{\nu }\frac{\partial }{\partial \xi ^{\mu }})
\end{array}
\label{2.1}
\end{equation}
be a representation of the $\Re _{\ell ,\infty }$ algebra (\ref{1.1}) by
differential operators in a 5-dimensional commutative manifold $M_{5}=\{\xi
_{a}\}$ with metric $\eta _{aa}=(1,-1,-1,-1,\epsilon )$

{\LARGE Case }$\epsilon =-1$

Changing to polar coordinates in the $\left( \xi ^{1},\xi ^{4}\right) $
plane ($\xi ^{1}=r\cos \theta ,\xi ^{4}=r\sin \theta $) 
\begin{equation}
\begin{array}{lll}
p^{1} & = & -i\left( \cos \theta \frac{\partial }{\partial r}-\frac{\sin
\theta }{r}\frac{\partial }{\partial \theta }\right)  \\ 
x^{1} & = & i\ell \frac{\partial }{\partial \theta }
\end{array}
\label{2.2}
\end{equation}

Eigenstates of the $x^{1}$ coordinate, with eigenvalue $\alpha $, are 
\begin{equation}
\left| \alpha \right\rangle =C_{\alpha }\left( r\right) \exp \left( -\frac{i%
}{\ell }\alpha \theta \right)   \label{2.3}
\end{equation}
$\theta \in S^{1}$ and $C_{\alpha }\left( r\right) $ an arbitrary $L^{2}-$%
function of $r$. Single-valuedness requires $\alpha \in \ell \Bbb{Z}$. That
is, each space coordinate has a discrete spectrum.

The eigenstates of $p^{1}$ (with eigenvalue $k$) are 
\begin{equation}
\left| k\right\rangle =\exp \left( ikr\cos \theta \right)   \label{2.4}
\end{equation}
They have a wave function representation in the position basis 
\begin{eqnarray}
\left\langle \alpha |k\right\rangle  &=&\int_{0}^{\infty }dr\int_{-\pi
}^{\pi }d\theta C_{\alpha }^{*}\left( r\right) e^{i\left( \frac{\alpha }{%
\ell }\theta +kr\cos \theta \right) }  \label{2.5} \\
&=&2\pi \left( i\right) ^{\frac{\alpha }{\ell }}\int_{0}^{\infty
}drC_{\alpha }^{*}\left( r\right) J_{\frac{\alpha }{\ell }}\left( kr\right) 
\nonumber
\end{eqnarray}

To obtain the density of states one imposes periodic boundary conditions in
a box of size $L$, leading to 
\begin{equation}
J_{0}\left( kr\right) =\left( i\right) ^{\frac{L}{\ell }}J_{\frac{L}{\ell }%
}\left( kr\right)   \label{2.6}
\end{equation}
For large $k$, using the asymptotic expansion for Bessel functions, Eq.(\ref
{2.6}) leads to 
\begin{equation}
\sqrt{\frac{2}{kr}}\left\{ \cos \left( kr-\frac{\pi }{4}\right) -\left(
i\right) ^{\frac{L}{\ell }}\cos \left( kr-\frac{L}{2\ell }\pi -\frac{\pi }{4}%
\right) +O\left( \left| kr\right| ^{-1}\right) \right\} =0  \label{2.7}
\end{equation}
Asymptotically, this is satisfied both for $\frac{L}{\ell }=2n,$ $n\in \Bbb{Z%
}$ and odd or $\frac{L}{\ell }=4n,$ $n\in \Bbb{Z}$.  Therefore, for very
large $k$, no restrictions are put on the $k$ values. It means that the
phase volume required for any new $k$ state shrinks as $k$ becomes large.
The density of states diverges for large $k$.

{\LARGE Case }$\epsilon =+1$

With hyperbolic coordinates $\left( \xi ^{1}=r\sinh \mu ,\xi ^{4}=r\cosh \mu
\right) $ in the $\left( \xi ^{1},\xi ^{4}\right) $ plane, 
\begin{equation}
\begin{array}{lll}
p^{1} & = & i\left( \sinh \mu \frac{\partial }{\partial r}-\frac{\cosh \mu }{%
r}\frac{\partial }{\partial \mu }\right)  \\ 
x^{1} & = & i\ell \frac{\partial }{\partial \mu }
\end{array}
\label{2.8}
\end{equation}

The eigenstates of the $x^{1}$ coordinate, with eigenvalue $\alpha $, are 
\begin{equation}
\left| \alpha \right\rangle =C_{\alpha }\left( r\right) \exp \left( -\frac{i%
}{\ell }\alpha \mu \right)   \label{2.9}
\end{equation}
Because $\mu \in \Bbb{R}$, in this case the space coordinates have
continuous spectrum. It is the time coordinate that has discrete spectrum.

The eigenstates of $p^{1}$ are 
\begin{equation}
\left| k\right\rangle =\exp \left( ikr\sinh \mu \right)   \label{2.10}
\end{equation}
with a wave function representation in the position basis 
\begin{eqnarray}
\left\langle \alpha |k\right\rangle  &=&\int_{0}^{\infty }dr\int_{-\infty
}^{\infty }d\mu C_{\alpha }^{*}\left( r\right) e^{i\left( \frac{\alpha }{%
\ell }\mu +kr\sinh \mu \right) }  \label{2.11} \\
&=&2\int_{0}^{\infty }drC_{\alpha }^{*}\left( r\right) K_{i\frac{\alpha }{%
\ell }}\left( kr\right) \left( \cosh \left( \frac{\alpha \pi }{2\ell }%
\right) -\sinh \left( \frac{\alpha \pi }{2\ell }\right) \right)   \nonumber
\end{eqnarray}

To obtain the density of states one imposes periodic boundary conditions in
a box of size $L$, leading to 
\begin{equation}
K_{0}\left( kr\right) =K_{i\frac{L}{\ell }}\left( kr\right) \left( \cosh
\left( \frac{L\pi }{2\ell }\right) -\sinh \left( \frac{L\pi }{2\ell }\right)
\right)   \label{2.12}
\end{equation}
Using the (large $z$) asymptotic expansion 
\begin{equation}
K_{\nu }\left( z\right) =\sqrt{\frac{\pi }{2z}}e^{-z}\left( 1+\frac{4\nu
^{2}-1}{8z}+O\left( z^{-2}\right) \right)   \label{2.13}
\end{equation}
leads for large $k$ to 
\begin{equation}
1-\cosh \left( \frac{L\pi }{2\ell }\right) +\sinh \left( \frac{L\pi }{2\ell }%
\right) +O\left( \left( kr\right) ^{-1}\right) =0  \label{2.14}
\end{equation}
This cannot be satisfied in $k\rightarrow \infty $ limit. It means that the
density of states vanishes for large $k$.

For arbitrary values of $k$ the exact density of states may be obtained from
Eqs.(\ref{2.6}) or (\ref{2.12}). However, to obtain a simpler, approximate,
form for the density of states it is convenient to use the representation of
a subalgebra. Namely, for the subalgebra $\left\{ x^{i},p^{i},\Im \right\} $
($i$ fixed $=1,2$ or $3$) one may use 
\begin{equation}
\begin{array}{lll}
x^{i} & = & x \\ 
p^{i} & = & \frac{1}{\ell }\sin \left( \frac{\ell }{i}\frac{d}{dx}\right) 
\\ 
\Im  & = & \cos \left( \frac{\ell }{i}\frac{d}{dx}\right) 
\end{array}
\label{2.15}
\end{equation}
for the $\epsilon =-1$ case and 
\begin{equation}
\begin{array}{lll}
x^{i} & = & x \\ 
p^{i} & = & \frac{1}{\ell }\sinh \left( \frac{\ell }{i}\frac{d}{dx}\right) 
\\ 
\Im  & = & \cosh \left( \frac{\ell }{i}\frac{d}{dx}\right) 
\end{array}
\label{2.16}
\end{equation}
for the $\epsilon =+1$ case.

The states 
\begin{equation}
\left| p\right\rangle =\exp \left( ikx\right)   \label{2.17}
\end{equation}
are eigenstates of $p^{i}$ corresponding to the eigenvalues 
\begin{equation}
\begin{array}{llll}
p\left( k\right)  & = & \frac{1}{\ell }\sin \left( k\ell \right)  & \text{%
for }\epsilon =-1 \\ 
p\left( k\right)  & = & \frac{1}{\ell }\sinh \left( k\ell \right)  & \text{%
for }\epsilon =+1
\end{array}
\label{2.18}
\end{equation}
Periodic boundary conditions for $\left| p\right\rangle $ on a box of size $L
$ implies 
\begin{equation}
k=\frac{2\pi }{L}n\hspace{1.2cm}n\in \Bbb{Z}  \label{2.19}
\end{equation}
From $dp=\frac{dp}{dn}dn$ one obtains for the density of states 
\begin{equation}
\begin{array}{llll}
dn & = & \frac{L}{2\pi }\frac{dp}{\sqrt{1-\ell ^{2}p^{2}}} & \text{for }%
\epsilon =-1 \\ 
dn & = & \frac{L}{2\pi }\frac{dp}{\sqrt{1+\ell ^{2}p^{2}}} & \text{for }%
\epsilon =+1
\end{array}
\label{2.20}
\end{equation}
The density of states vanishes when $p\rightarrow \infty $ in the $\epsilon
=+1$ case and for $\epsilon =-1$ it diverges at $p=\frac{1}{\ell }$ (which
is the upper bound of the momentum in this case). This result is consistent
with what has been obtained from the asymptotic form of Eqs.(\ref{2.6}) and (%
\ref{2.12}). However, the density of states in Eqs.(\ref{2.20}) is not exact
because it is derived from a subalgebra representation, which cannot be
lifted in this simple form to a full representation of the algebra .

The modification of the phase-space volume implies corresponding
modifications of the cross sections. As an example, to be used in the
calculations of the next section, consider the reaction 
\begin{equation}
\gamma +p\rightarrow \pi +N  \label{2.20a}
\end{equation}
at high incident proton energy.

Here and in Section 3, simple letters are used to denote quantities in the
laboratory frame, primed letters for the rest frame of the incident proton
and starred letters for the center of mass. Using (\ref{2.20}), the modified
part of the phase-space integration in the cross section is 
\begin{equation}
I\left( \ell \right) =\int \int \frac{k_{\pi }^{2}dk_{\pi }}{\omega _{\pi }%
\sqrt{1+\epsilon \ell ^{2}k_{\pi }^{2}}}\frac{p_{N}^{2}dp_{N}}{E_{N}\sqrt{%
1+\epsilon \ell ^{2}p_{N}^{2}}}d\Omega _{\pi }d\Omega _{N}\delta ^{4}\left(
p_{\gamma }+p_{p}-k_{\pi }-p_{N}\right)   \label{2.21}
\end{equation}
At high energies, with quantities in the rest frame of the incident proton,
one obtains 
\begin{equation}
I\left( \ell \right) \sim \int_{0}^{\omega _{\gamma }^{\prime }}\frac{%
k^{\prime }\left( \omega _{\gamma }^{\prime }-k^{\prime }\right) dk^{\prime }%
}{\sqrt{1+\epsilon \ell ^{2}k^{\prime 2}}\sqrt{1+\epsilon \ell ^{2}\left(
\omega _{\gamma }^{\prime }-k^{\prime }\right) ^{2}}}  \label{2.22}
\end{equation}
Changing variables and dividing by $I\left( 0\right) $ one obtains the
following suppression ($\epsilon =+1$) or enhancing ($\epsilon =-1$)
function 
\begin{equation}
g\left( \alpha ,\epsilon \right) =\frac{I\left( \ell \right) }{I\left(
0\right) }\simeq 6\int_{0}^{1}\frac{x\left( 1-x\right) dx}{\sqrt{1+\epsilon
\alpha x^{2}}\sqrt{1+\epsilon \alpha \left( 1-x\right) ^{2}}}  \label{2.23}
\end{equation}
with $\alpha =\omega _{\gamma }^{\prime 2}\ell ^{2}$. Fig.1 is a plot of
this function in the $\epsilon =+1$ (suppression) case.

\begin{figure}[htb]
\begin{center}
\psfig{figure=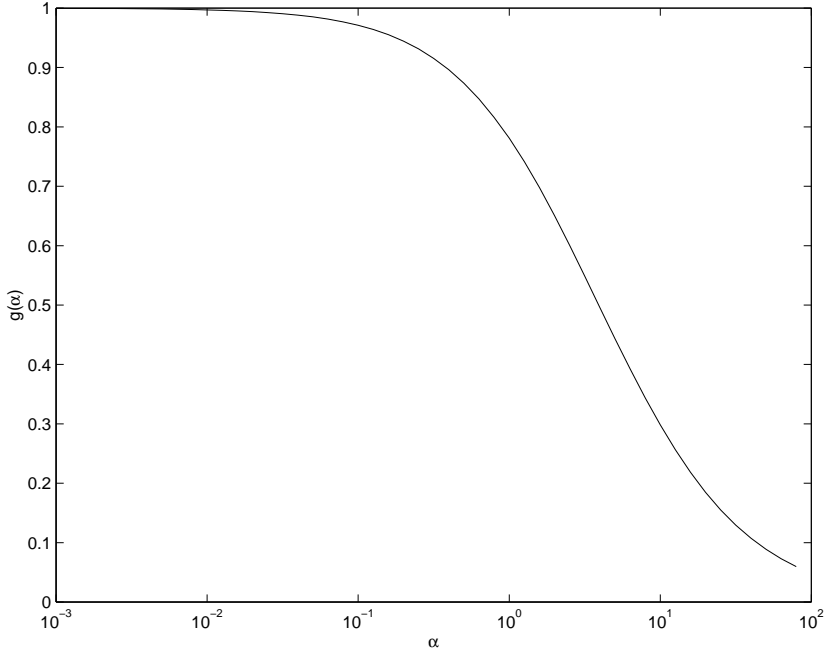,width=11truecm}
\end{center}
\caption{The phase-space suppression function ($\varepsilon =+1$ case)}
\end{figure}

\section{The GZK sphere}

In the sixties, Greisen \cite{Greisen}, Zatsepin and Kuz'min \cite{Zatsepin}
have shown that the cosmic microwave background radiation should make the
Universe opaque to protons of energies $\gtrsim $ $10^{20}\,$eV. At these
energies the thermal photons are sufficiently blue-shifted in the proton
rest frame to excite baryon resonances and drain the proton's energy via
pion production. This led to the notion of \textit{GZK-sphere} as being the
sphere within which a source has to lie to supply us with protons at $%
10^{20}\,$eV. Later, more accurate calculations, using state-of-the-art
particle physics data, placed the energy limit of cosmic (not arising from
local sources) protons at around $5.10^{19}\,$eV. That is, if the proton
sources are at cosmological distances ($\gtrsim $ $100$ Mpc), the observed
spectrum should display a (GZK) cutoff around this energy. A similar limit
applies to the nuclei component of the cosmic ray flux.

This situation was upset by the detection of a number of events above $%
10^{20}\,$eV without any plausible local sources \cite{Bird} \cite{Takeda1} 
\cite{Takeda2}. Discrepancies between the fluxes measured by different
groups \cite{Abu1} \cite{Abu2} and analysis of the combined data \cite
{Bahcall} do not yet allow for a clear-cut statement as to whether the GZK\
cutoff is indeed violated, a question that will hopefully be clarified by
the forthcoming Auger observatory. Meanwhile a number of possible
explanations for the violation of the GZK\ cutoff has appeared on the
literature (for a review see \cite{Anchordoqui1}). Here I analyze the effect
of the space-time non-commutativity on the calculation of the GZK cutoff
and, when (and if) such cutoff is confirmed, what inferences can be taken
concerning the value of $\ell $ and the sign $\epsilon $.

Simple letters are used to denote quantities in the lab (earth) frame,
primed letters for the rest frame of the proton and starred letters for the
center of mass. The fractional energy loss due to interactions with the
cosmic background radiation (at zero redshift) is given by the integral of
the nucleon energy loss per collision multiplied by the probability per unit
time for a nucleon-photon collision in an isotropic gas of photons at
temperature $T=2.7^{o}K$. Therefore the lifetime of a cosmic ray of energy $E
$ is \cite{Stecker1}, $\left( \hbar =c=1\right) $ 
\begin{equation}
\tau _{0}\left( E\right) =2\Gamma ^{2}\pi ^{2}\left\{ \sum_{j}\int_{\omega
_{jth}^{^{\prime }}/2\Gamma }^{\infty }\frac{d\omega }{e^{\omega /kT}-1}%
\int_{\omega _{jth}^{^{\prime }}}^{2\Gamma \omega }d\omega ^{^{\prime
}}\omega ^{^{\prime }}\sigma _{j}\left( \omega ^{^{\prime }}\right)
K_{j}\left( \omega ^{^{\prime }}\right) \right\} ^{-1}  \label{3.1}
\end{equation}
where $\omega ^{^{\prime }}$ is the photon energy in the nucleon rest frame
and the inelasticity $K_{j}$ is the average energy lost by the photon for
the channel $j$ with threshold $\omega _{jth}^{^{\prime }}$. $\sigma
_{j}\left( \omega ^{^{\prime }}\right) $ is the total cross section of the $j
$-th interaction channel and $\Gamma $ the Lorentz factor of the nucleon $%
\left( \Gamma =\frac{E}{m_{p}}\right) $.

In (\ref{3.1}) one may change the order of integration
\[
\int_{\omega _{th}^{^{\prime }}/2\Gamma }^{\infty }d\omega \int_{\omega
_{th}^{^{\prime }}}^{2\Gamma \omega }d\omega ^{\prime }\rightarrow
\int_{\omega _{th}^{^{\prime }}}^{\infty }d\omega ^{\prime }\int_{\omega
^{^{\prime }}/2\Gamma }^{\infty }d\omega 
\]
and compute one of the integrals.. To obtain the cosmic ray lifetime $\tau
_{\ell }\left( E\right) $ in the non-commutative case, one multiplies the
cross section by the suppression factor $g\left( \omega ^{\prime },\epsilon
\right) $ (Eq.(\ref{2.23})). Finally, changing variables
\[
\omega ^{\prime }\rightarrow y=e^{-\omega ^{\prime }/\left( 2\Gamma
kT\right) }
\]
one obtains the following ratio for each channel contribution 
\begin{equation}
r_{g}=\frac{\tau _{\ell }\left( E\right) }{\tau _{0}\left( E\right) }=\frac{%
\int_{e^{\frac{\omega _{th}^{^{\prime }}}{\beta }}}^{0}\frac{dy}{y}\ln
\left( 1-y\right) \ln y\sigma \left( \beta \ln y\right) K\left( \beta \ln
y\right) }{\int_{e^{\frac{\omega _{th}^{^{\prime }}}{\beta }}}^{0}\frac{dy}{y%
}\ln \left( 1-y\right) \ln y\sigma \left( \beta \ln y\right) K\left( \beta
\ln y\right) g\left( \beta ^{2}\ell ^{2}\ln ^{2}y,\epsilon \right) }
\label{3.2}
\end{equation}
with $\beta =-2\Gamma kT$.

This ratio was estimated for the single pion reaction (\ref{2.20a}) using $%
\omega _{th}^{\prime }=145$ MeV,  
\[
K\left( \omega ^{\prime }\right) =\frac{1}{2}\left( 1+\frac{m_{\pi
}^{2}-m_{N}^{2}}{m_{p}^{2}+2m_{p}\omega ^{\prime }}\right) 
\]
and the following parametrization \cite{data} for the cross section 
\[
\sigma \left( \omega ^{\prime }\right) =A+B\ln ^{2}\left( \omega ^{\prime
}\right) +C\ln \left( \omega ^{\prime }\right) 
\]
with $A=0.147,B=0.0022,C=-0.017$ , $\omega ^{\prime }$ in Gev's. This is a
parametrization for the  $\gamma p$ total cross section in the range $3$GeV$%
<\omega ^{\prime }<183$GeV. Of course, to compute the absolute value of $%
\tau _{\ell }\left( E\right) $ this would not be appropriate. Instead, due
account should be taken of all the resonances contributions. However for the
ratio $r_{g}$ it gives, at least, qualitative information on the order of
magnitude of the effect.

\begin{figure}[htb]
\begin{center}
\psfig{figure=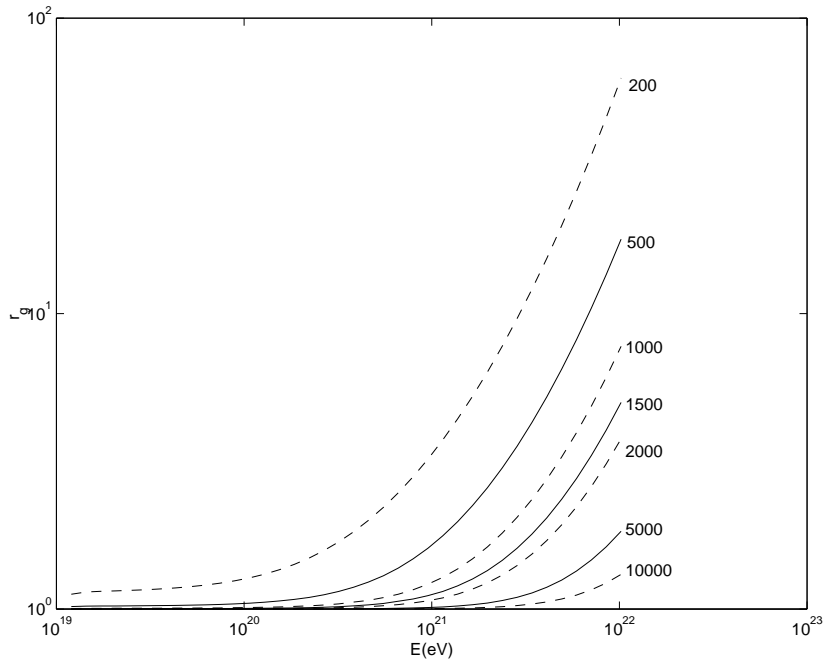,width=11truecm}
\end{center}
\caption{Cosmic ray lifetime increasing factors for $1/\ell $ in the range $%
200$ to $10000$ MeV ($\varepsilon =+1$ case)}
\end{figure}

In Fig.2 the results for $r_{g}$ are shown for $\epsilon =+1$ and $1/\ell $
in the range $200$ to $10000$ Mev, that is, $\ell $ in the range $0.98-0.0197
$ Fermi or $329-6.58\times 10^{-26}$ seconds.

To estimate the effect that this lifetime extending factors have on the
energy attenuation of cosmic rays on route to earth, I have used the $\left(
dD/dE\right) _{\infty }$ values found in \cite{Anchordoqui2} for a $10^{22}$
eV nucleon and computed the integration
\begin{equation}
D\left( E\right) =D_{0}\left( E_{0}\right) +\int_{E_{0}}^{E}r_{g}\left(
E\right) \left( dD/dE\right) _{\infty }dE  \label{3.3}
\end{equation}
The results are shown in Fig. 3. One sees that whereas the value of the GZK
cutoff\ is not much changed, the radius of the GZK sphere is increased
allowing for nucleons from distances beyond $100$ Mpc to reach earth at
energies above $5.10^{19}$ eV. 

\begin{figure}[htb]
\begin{center}
\psfig{figure=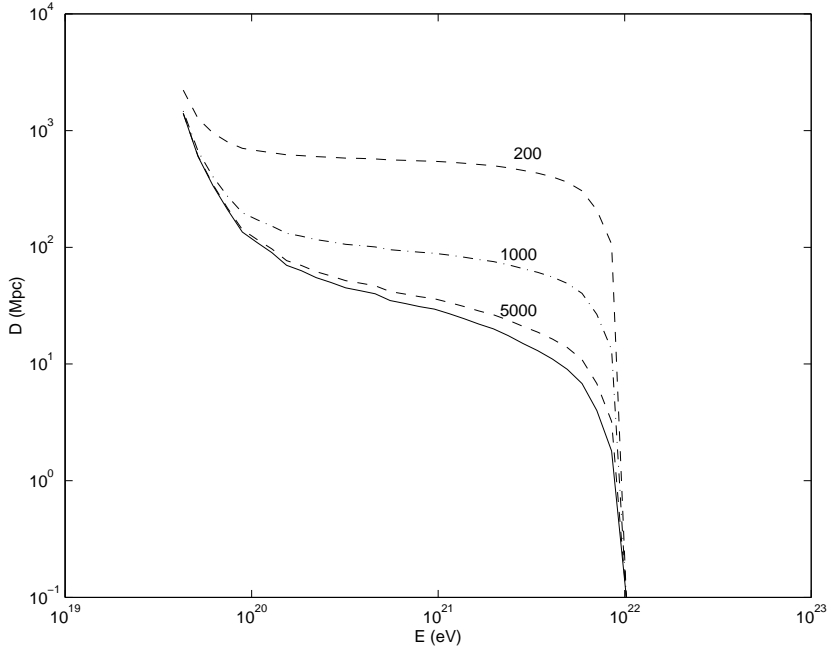,width=11truecm}
\end{center}
\caption{Energy attenuation of a $10^{22}$ eV nucleon in route to earth: $%
1/\ell =200,1000,5000$ MeV compared with the $\ell =0$ case}
\end{figure}

If the observation of the ultra high energy cosmic rays is indeed a
manifestation of the non-commutative structure two conclusions may be taken:

- First, that the sign $\epsilon $ is $+1$, that is, space is continuous and
time discrete.

- Second, that for the effect to be significant at current cosmic ray
energies, the time quantum must be $\gtrsim 10^{-25}$ seconds, much larger
than Planck scale times.

\section{Corrections to the spectrum of the Coulomb problem}

Some experiments in atomic physics are now sensitive to small frequency
shifts below 1 mHz. With such sensitivity, non-commutative space-time
effects might be detected at low energies, especially if small energy shifts
have a qualitative impact. Here such a possibility is analyzed by looking at
the effect of the non-commutative algebra on the spectrum of the Coulomb
problem. Consider the Hamiltonian

\begin{equation}
H=-\frac{1}{2m}\Delta -\frac{e^{2}}{\left| \overrightarrow{x}\right| }
\label{4.1}
\end{equation}
and use, for the non-commutative coordinates and momenta, the representation
listed in the Appendix. Both cases ($\epsilon =-1$ and $\epsilon =+1$) will
be considered.

\fbox{$\epsilon =-1$}

From (\ref{A.3}) one obtains (setting $R=1$) 
\begin{equation}
\begin{array}{lll}
\left| \overrightarrow{p}\right| ^{2} & = & \frac{1}{\ell ^{2}}\sin
^{2}\theta _{3} \\ 
\left| \overrightarrow{x}\right| ^{2} & = & \ell ^{2}\left\{ L^{2}\cot
^{2}\theta _{3}-\frac{\partial ^{2}}{\partial \theta _{3}^{2}}-2\cot \theta
_{3}\frac{\partial }{\partial \theta _{3}}\right\} 
\end{array}
\label{4.2}
\end{equation}

Therefore 
\begin{equation}
H=\frac{1}{2m\ell ^{2}}\sin ^{2}\theta _{3}-\frac{e^{2}}{\ell }\left\{
L^{2}\cot ^{2}\theta _{3}-\frac{\partial ^{2}}{\partial \theta _{3}^{2}}%
-2\cot \theta _{3}\frac{\partial }{\partial \theta _{3}}\right\} ^{-\frac{1}{%
2}}  \label{4.3}
\end{equation}

For small $\ell $, (small $\theta _{3}$) one obtains 
\begin{equation}
H\simeq \frac{1}{2m}\left\{ \frac{\theta _{3}^{2}}{\ell ^{2}}-\frac{\ell ^{2}%
}{3}\left( \frac{\theta _{3}}{\ell }\right) ^{4}\right\} -e^{2}\left\{
L^{2}\left( \frac{\ell ^{2}}{\theta _{3}^{2}}-\frac{2\ell ^{2}}{3}\right)
-\ell ^{2}\frac{\partial ^{2}}{\partial \theta _{3}^{2}}-2\ell ^{2}\left( 
\frac{1}{\theta _{3}}-\frac{\theta _{3}}{3}\right) \frac{\partial }{\partial
\theta _{3}}\right\} ^{-\frac{1}{2}}  \label{4.4}
\end{equation}
In this approximation $p\simeq \frac{\theta _{3}}{\ell }$, therefore 
\[
H\simeq \frac{1}{2m}\left\{ p^{2}-\frac{\ell ^{2}}{3}p^{4}\right\}
-e^{2}\left\{ L^{2}\left( \frac{1}{p^{2}}-\frac{2\ell ^{2}}{3}\right) -\frac{%
\partial ^{2}}{\partial p^{2}}-2\left( \frac{1}{p}-\frac{\ell ^{2}}{3}%
p\right) \frac{\partial }{\partial p}\right\} ^{-\frac{1}{2}}
\]
which may be rewritten 
\begin{equation}
H\simeq \frac{1}{2m}p^{2}-\frac{e^{2}}{\left( \nabla _{p}^{2}\right) ^{\frac{%
1}{2}}}+\ell ^{2}\left\{ -\frac{1}{6m}p^{4}-\frac{e^{2}\left( L^{2}-p\frac{%
\partial }{\partial p}\right) }{3\left( \nabla _{p}^{2}\right) ^{\frac{3}{2}}%
}\right\}   \label{4.5}
\end{equation}
with $\nabla _{p}^{2}=\frac{L^{2}}{p^{2}}-\frac{\partial ^{2}}{\partial p^{2}%
}-\frac{2}{p}\frac{\partial }{\partial p}$

Using the Fourier transform $f\left( x\right) =\int e^{ip.x}F\left( p\right)
d^{3}p$ and the relations 
\begin{equation}
\begin{array}{lll}
\int e^{ip.x}p^{2}F\left( p\right) d^{3}p & = & -\nabla _{x}^{2}f\left(
x\right)  \\ 
\int e^{ip.x}\nabla _{p}^{2}F\left( p\right) d^{3}p & = & -x^{2}f\left(
x\right)  \\ 
\int e^{ip.x}p\frac{\partial }{\partial p}F\left( p\right) d^{3}p & = & 
\left( -r\frac{\partial }{\partial r}-3\right) f\left( x\right) 
\end{array}
\label{4.6}
\end{equation}
one obtains a configuration space representation of Eq.(\ref{4.5}), namely 
\begin{equation}
H\simeq -\frac{\nabla _{x}^{2}}{2m}-\frac{e^{2}}{\left| x\right| }-\ell
^{2}\left\{ \frac{1}{6m}\nabla _{x}^{4}+\frac{e^{2}}{3}\frac{L^{2}+r\frac{%
\partial }{\partial r}+3}{r^{3}}\right\}   \label{4.7}
\end{equation}
The first two terms are the usual Coulomb Hamiltonian and the third is the
order $\ell ^{2}$ correction arising from the non-commutative structure. 
\begin{equation}
\left\langle n^{\prime }L^{^{\prime }}M^{^{\prime }}\left| H\right|
nLM\right\rangle \simeq \delta _{LL^{^{\prime }}}\delta _{MM^{^{\prime
}}}\left\{ E_{n}\delta _{n^{\prime },n}+\ell ^{2}\left\langle -\frac{1}{6m}%
\nabla _{x}^{4}-\frac{e^{2}}{3}\frac{L\left( L+1\right) +r\frac{\partial }{%
\partial r}+3}{r^{3}}\right\rangle _{n^{\prime },n}\right\}   \label{4.8}
\end{equation}
where $r=\left| \overrightarrow{x}\right| $.

\fbox{$\epsilon =+1$}

For the $\epsilon =+1$ case one uses the same representation with the
replacements $x^{\nu }\rightarrow ix^{\nu },p^{\nu }\rightarrow -ip^{\nu
},\theta _{3}\rightarrow i\mu $, to obtain
\begin{equation}
\begin{array}{lll}
\left| \overrightarrow{p}\right| ^{2} & = & \frac{1}{\ell ^{2}}\sinh ^{2}\mu 
\\ 
\left| \overrightarrow{x}\right| ^{2} & = & \ell ^{2}\left\{ L^{2}\coth
^{2}\mu -\frac{\partial ^{2}}{\partial \mu ^{2}}-2\coth \mu \frac{\partial }{%
\partial \mu }\right\} 
\end{array}
\label{4.9}
\end{equation}
Then
\begin{equation}
H=\frac{1}{2m\ell ^{2}}\sinh ^{2}\mu -\frac{e^{2}}{\ell }\left\{ L^{2}\coth
^{2}\mu -\frac{\partial ^{2}}{\partial \mu ^{2}}-2\coth \mu \frac{\partial }{%
\partial \mu }\right\} ^{-\frac{1}{2}}  \label{4.10}
\end{equation}
and for small $\mu $%
\[
H\simeq \frac{1}{2m}\left\{ \frac{\mu ^{2}}{\ell ^{2}}+\frac{\ell ^{2}}{3}%
\left( \frac{\mu }{\ell }\right) ^{4}\right\} -e^{2}\left\{ L^{2}\left( 
\frac{\ell ^{2}}{\mu ^{2}}+\frac{2\ell ^{2}}{3}\right) -\ell ^{2}\frac{%
\partial ^{2}}{\partial \mu ^{2}}-2\ell ^{2}\left( \frac{1}{\mu }+\frac{\mu 
}{3}\right) \frac{\partial }{\partial \mu }\right\} ^{-\frac{1}{2}}
\]
\[
H\simeq \frac{1}{2m}\left\{ p^{2}+\frac{\ell ^{2}}{3}p^{4}\right\}
-e^{2}\left\{ L^{2}\left( \frac{1}{p^{2}}+\frac{2\ell ^{2}}{3}\right) -\frac{%
\partial ^{2}}{\partial p^{2}}-2\left( \frac{1}{p}+\frac{\ell ^{2}}{3}%
p\right) \frac{\partial }{\partial p}\right\} ^{-\frac{1}{2}}
\]
\[
H\simeq \frac{1}{2m}p^{2}-\frac{e^{2}}{\left( \nabla _{p}^{2}\right) ^{\frac{%
1}{2}}}+\ell ^{2}\left\{ \frac{1}{6m}p^{4}+\frac{e^{2}\left( L^{2}-p\frac{%
\partial }{\partial p}\right) }{3\left( \nabla _{p}^{2}\right) ^{\frac{3}{2}}%
}\right\} 
\]
\begin{equation}
H\simeq -\frac{\nabla _{x}^{2}}{2m}-\frac{e^{2}}{\left| x\right| }+\ell
^{2}\left\{ \frac{1}{6m}\nabla _{x}^{4}+\frac{e^{2}}{3}\frac{L^{2}+r\frac{%
\partial }{\partial r}+3}{r^{3}}\right\}   \label{4.11}
\end{equation}
the conclusion being that for the $\epsilon =+1$ case the order $\ell ^{2}$
correction differs from the $\epsilon =-1$ case by a sign change.

\section{Conclusions}

1) A non-commutative space-time structure and two constants of Nature $\ell $
and $\epsilon $ emerge as natural consequences of deformation-theory and
stability of the fundamental physical theories. Among other effects, this
structure implies a modification of phase-space volume which, in particular,
has a bearing on the calculation of the GZK sphere. Lorentz invariance is
preserved.

2) Phase-space suppression effects occur only in the $\epsilon =+1$ case. In
this case the time coordinate has a discrete spectrum and space coordinates
are continuous.

3) In addition to changing the cross sections of elementary processes,
phase-space counting rules have statistical mechanics consequences which
might have had a relevant effect at the first stages of the Universe
evolution.

4) Phase-space volume modifications, time and space coordinates spectra and
modifications of the uncertainty relations are consequences of the
non-commutative space-time structure which depend only on its algebraic
structure. In this sense they are very robust and provide unambiguous tests
of the theory. Other consequences might depend on the particular geometric
construction that is built on top of the algebraic structure. For example,
for a particular geometrical construction \cite{Vilela3} the existence of
additional components on gauge fields is an intriguing consequence.

{\LARGE Appendix A}

For specific calculations it is convenient to use a representation of the
space-time algebra ($\varepsilon =-1$ case) in the space of functions on the
upper sheet of the cone $C^{4}$, with coordinates 
\begin{equation}
\begin{array}{lll}
\xi _{1} & = & R\sin \theta _{3}\sin \theta _{2}\sin \theta _{1} \\ 
\xi _{2} & = & R\sin \theta _{3}\sin \theta _{2}\cos \theta _{1} \\ 
\xi _{3} & = & R\sin \theta _{3}\cos \theta _{2} \\ 
\xi _{4} & = & R\cos \theta _{3} \\ 
\xi _{5} & = & R
\end{array}
\label{A.1}
\end{equation}
the invariant measure for which the functions are square-integrable being 
\begin{equation}
d\nu (R,\theta _{i})=R^{2}\sin ^{2}\theta _{3}\sin \theta _{2}dRd\theta
_{1}d\theta _{2}d\theta _{3}  \label{A.2}
\end{equation}

On these functions the operators of $\Re _{\ell ,\infty }$ act as follows 
\begin{equation}
\begin{array}{lll}
\ell p^{0} & = & R \\ 
\Im & = & R\cos \theta _{3} \\ 
\ell p^{1} & = & R\sin \theta _{3}\cos \theta _{2} \\ 
\ell p^{2} & = & R\sin \theta _{3}\sin \theta _{2}\cos \theta _{1} \\ 
\ell p^{3} & = & R\sin \theta _{3}\sin \theta _{2}\sin \theta _{1} \\ 
M^{23} & = & -i\frac{\partial }{\partial \theta _{1}} \\ 
M^{12} & = & -i\left( \cos \theta _{1}\frac{\partial }{\partial \theta _{2}}%
-\sin \theta _{1}\cot \theta _{2}\frac{\partial }{\partial \theta _{1}}%
\right) \\ 
M^{31} & = & i\left( \sin \theta _{1}\frac{\partial }{\partial \theta _{2}}%
+\cos \theta _{1}\cot \theta _{2}\frac{\partial }{\partial \theta _{1}}%
\right) \\ 
\frac{x^{0}}{\ell } & = & -i\left( -\sin \theta _{3}\frac{\partial }{%
\partial \theta _{3}}+R\cos \theta _{3}\frac{\partial }{\partial R}\right)
\\ 
\frac{x^{1}}{\ell } & = & i\left( \cos \theta _{2}\frac{\partial }{\partial
\theta _{3}}-\sin \theta _{2}\cot \theta _{3}\frac{\partial }{\partial
\theta _{2}}\right) \\ 
\frac{x^{2}}{\ell } & = & i\left( \cos \theta _{1}\sin \theta _{2}\frac{%
\partial }{\partial \theta _{3}}+\cos \theta _{1}\cos \theta _{2}\cot \theta
_{3}\frac{\partial }{\partial \theta _{2}}-\frac{\sin \theta _{1}}{\sin
\theta _{2}}\cot \theta _{3}\frac{\partial }{\partial \theta _{1}}\right) \\ 
\frac{x^{3}}{\ell } & = & i\left( \sin \theta _{1}\sin \theta _{2}\frac{%
\partial }{\partial \theta _{3}}+\sin \theta _{1}\cos \theta _{2}\cot \theta
_{3}\frac{\partial }{\partial \theta _{2}}+\frac{\cos \theta _{1}}{\sin
\theta _{2}}\cot \theta _{3}\frac{\partial }{\partial \theta _{1}}\right) \\ 
M^{01} & = & i\left( \frac{\sin \theta _{2}}{\sin \theta _{3}}\frac{\partial 
}{\partial \theta _{2}}-\cos \theta _{2}\cos \theta _{3}\frac{\partial }{%
\partial \theta _{3}}-R\cos \theta _{2}\sin \theta _{3}\frac{\partial }{%
\partial R}\right) \\ 
M^{02} & = & -i\left( 
\begin{array}{c}
\frac{\cos \theta _{1}\cos \theta _{2}}{\sin \theta _{3}}\frac{\partial }{%
\partial \theta _{2}}-\frac{\sin \theta _{1}}{\sin \theta _{2}\sin \theta
_{3}}\frac{\partial }{\partial \theta _{1}}+\cos \theta _{1}\sin \theta
_{2}\cos \theta _{3}\frac{\partial }{\partial \theta _{3}}+ \\ 
R\cos \theta _{1}\sin \theta _{2}\sin \theta _{3}\frac{\partial }{\partial R}
\end{array}
\right) \\ 
M^{03} & = & -i\left( 
\begin{array}{c}
\frac{\sin \theta _{1}\cos \theta _{2}}{\sin \theta _{3}}\frac{\partial }{%
\partial \theta _{2}}+\frac{\cos \theta _{1}}{\sin \theta _{2}\sin \theta
_{3}}\frac{\partial }{\partial \theta _{1}}+\sin \theta _{1}\sin \theta
_{2}\cos \theta _{3}\frac{\partial }{\partial \theta _{3}}+ \\ 
R\sin \theta _{1}\sin \theta _{2}\sin \theta _{3}\frac{\partial }{\partial R}
\end{array}
\right)
\end{array}
\label{A.3}
\end{equation}

For the $\varepsilon =+1$ case, one may work out a similar representation on
the $C^{3,1}$ cone with coordinates 
\begin{equation}
\begin{array}{lll}
\zeta _{1} & = & R\cosh \beta \cos \psi _{0} \\ 
\zeta _{2} & = & R\cosh \beta \sin \psi _{0} \\ 
\zeta _{3} & = & R\sinh \beta \sin \psi _{1} \\ 
\zeta _{4} & = & R\sinh \beta \cos \psi _{1} \\ 
\zeta _{5} & = & R
\end{array}
\label{A.4}
\end{equation}
Alternatively we may use the above representation multiplying $x^{\mu }$ by $%
i$, $p^{\mu }$ by $-i$ and replacing $\theta _{3}$ by $i\mu $. It is easily
seen from (\ref{1.1}) that the correct commutation relations, for the $%
\varepsilon =+1$ case, are obtained.

\end{document}